\documentclass{jltp}

\usepackage{graphicx}
\usepackage{epsfig, graphicx, euscript}

\title{Slow Transient Processes in the Second Sound Resonator}

\author{ Sergey  K. Nemirovskii, Krotov S. V., Sorokin A.L.}

\address{Institute of Thermophysics SB RAS, 630090 Novosibirsk,
Russia}

\runninghead{ S. Nemirovskii}{Slow Transient Processes}

\begin{document}

\maketitle

\begin{abstract}
The Hydrodynamics of Superfluid Turbulence (HST) describes the
flows (or counterflows) of HeII in the presence of a chaotic set
of vortex filaments. The HST equations govern both a slow
variation of the hydrodynamic variables due to dissipation related
to the vortex tangle and fast processes of the first and second
sound propagation. This circumstance prevents effective numerical
simulations of the problems of unsteady heat transfer in HeII. By
virtue of a pertinent multi-scale perturbation analysis we show
how one can eliminate the fast processes to derive the evolution
equation for the slow processes only. We then demonstrate that the
long-term evolution of a transient heat load of moderate intensity
obeys the nonlinear heat conductivity equation. The second example
of the methods developed is investigation of unsteady processes in
the second sound resonator. The latter is frequently used for
study of nonstationary behavior of vortex tangle, just by
monitoring of the quality factor behavior. This procedure however
is wrong when characteristic times of processes are comparable (or
smaller) than the time constant of resonator. We show how to
extract the correct information on the vortex line density (VLD)
dynamics with use of procedure we developed.

PACS numbers: 47.32.Cc, 47.37.+q, 67.40.Vs., 05.10.Gg.
\end{abstract}
\maketitle

\section{ INTRODUCTION AND SCIENTIFIC BACKGROUND}
It is appreciated that in most cases the flow (or counterflow) of
HeII occurs in the presence of a chaotic set of vortex filaments,
the so called superfluid turbulence. To describe various problems
both for applications and for experiments it is necessary to use
equations of HST. These equations unify classical equations of
hydrodynamics of HeII and the famous Vinen equation describing the
dynamics of the vortex tangle. The HST was widely used to describe
various hydrodynamics as well as thermal phenomena in HeII in
presence of vortex tangle (see the review paper \cite{NF}).

The set of HST equations is extremely cumbersome, therefore, there
is no wonder that to achieve quantitative results one is bound to
turn to numerical methods. Yet, numerical simulation of
nonstationary flows of HeII faces one serious obstacle. The point
here is that slow variation of hydrodynamic variables is
accompanied by the fast processes related to propagation of the
first and second sounds. If one is interested in the slow
evolution, particular details of sound propagation become
completely irrelevant, yet requiring rather extensive numerical
resources.

Thus, it seems attractive to try to get rid of the fast modes by a
pure analytical procedure. In the present paper we realize
effective separation of the slow from the fast modes using
multi-time asymptotic perturbation techniques (See, for example,
book by Nayfeh \cite{Nayfeh}).  That is quite universal powereful
method, its application, however, depends on concrete statement of
the problem. In the present work we  consider two particular
examples. In the first example we follow how the initially
hyperbolic-type equations of HST (describing wave type of heat
transfer) lead to purely parabolic-type nonlinear heat transfer
equations. Another example of the method is the study of unsteady
processes in the second sound resonator.

\section{ THE NONLINEAR HEAT CONDUCTIVITY EQUATION}

Let us consider the problem of one-dimensional heat exchange in
the tube filled by HeII. We assume that  a heater is switched
instantly on and the thermal front starts out propagating into
undisturbed bulk of helium. Initially, of course, the propagation
of the heat should be realized by the second sound mechanism. But
gradually, due to quantum vorticity the second sound attenuates
and degenerates. The mechanism of the heat  exchange should
change. In order to ascertain both this new mechanism and the
correspondent law  we apply multi-time asymptotic perturbation
techniques.

The set of equations of HST  for dimensionless velocity of the
normal component $ V_{n}^{\prime }$ , dimensionless temperature
$T^{\prime }$ and dimensionless square root of the vortex line
density (VLD) $G=\sqrt{\mathcal{L}/ \mathcal{L_{\infty }}}$ ( for
details and notations see the review article of Nemirovskii and
Fiszdon \cite{NF}) is reduced to the form:

\begin{equation}
\frac{\partial V_{n}^{\prime }}{\partial t^{\prime
}}\;+\;\frac{\partial T^{\prime }}{\partial x^{\prime
}}\;=\;-\;Sh\;G^{2}V_{n}^{\prime },\;\;\;\;\;\;\;\;\frac{\partial
T^{\prime }}{\partial t^{\prime }}\;+\;\;\frac{\partial
V_{n}^{\prime }}{\partial x^{\prime }}\;=0, \label{V}
\end{equation}

\begin{equation}
\frac{\partial G}{\partial t^{\prime }}\;=\;\frac{\alpha
}{A(T){\rho }_{s}{ \rho }_{n}}\;(\frac{\alpha }{\beta
})\;\frac{\;Sh}{2}\;(G^{2}V_{n}^{\prime }\;-\;G^{3}),  \label{G}
\end{equation}
with the following dimensionless variables:
\begin{equation}
t\;=\;\frac{L}{c_{2}}t^{\prime }, \;\;x\;=\;Dx^{\prime },\;\;
V_{n}\;=\;V_{n0}V_{n}^{\prime },\;\; \;\;T\;=\;\frac{\sigma
V_{n0}}{{\sigma } _{T}c_{2}}T^{\prime
},\;\;\;\;\mathcal{L}\;=\;\mathcal{L}_{\infty }G^{2}. \label{DIM}
\end{equation}

It is easy to see  that the dynamics of the hydrodynamic variables
is specified by the dimensionless criterion the Strouhal number $
Sh\;=D/c_{2}{\tau }_{d}$, ($D$-length of channel, ${\tau }_{d}$
decrement of attenuation of counterflow due to vortices). The
Strouhal number is defined as the ratio of the counterflow
decrement to inverse time that takes the heat pulse to cross the
channel. In the case of heat load of moderate intensity, of order
of a few $W/cm^{2}$, the quantity $Sh$ is of order of
$10^{2}-10^{5}$, i.e. much greater than unity.

 For times of order $L/(c_{2}Sh)$ the heat pulse propagates
according to wave-like equation. Our goal now is to clarify what
will happen at later times. For that purpose we have to eliminate
the fast processes. According general ideas of multi-scale
asymptotic perturbation theory (see e.g.Nayfeh \cite{Nayfeh}) we
look for a solution to the set of the HST equations
(\ref{V})-(\ref{G}) in the form of an asymptotic series. Following
this method, we introduce different time scales. $t_{0}=t^{\prime
};\qquad \;t_{1}^{\prime }=\epsilon t^{\prime };\qquad
t_{2}^{\prime }=\epsilon ^{2}t^{\prime }\mathrm{\ \qquad
.......,}$ where $\epsilon =1/Sh<<1$. We look for a solution to
the set of the HST equations (\ref{V})-(\ref{G}) in the form of an
asymptotic series $ V_{n}^{\prime }=V_{0}^{\prime }(x^{\prime
},t_{0}^{\prime },t_{1}^{\prime },t_{2}^{\prime })+\epsilon
V_{1}^{\prime }(x^{\prime },t_{0}^{\prime },t_{1}^{\prime
},t_{2}^{\prime })+\epsilon ^{2}V_{2}^{\prime }(x^{\prime
},t_{0}^{\prime },t_{1}^{\prime },t_{2}^{\prime })+....$ and
similarly to $T^{\prime }$ and $G$. The simple rule takes place:
\begin{equation}
\frac{\partial }{\partial t^{\prime }}=\frac{\partial }{\partial
t_{0}^{\prime }}+\epsilon \frac{\partial }{\partial t_{1}^{\prime
}} +\epsilon ^{2}\frac{\partial }{\partial t_{2}^{\prime }}
\end{equation}
The next step in study of the slow evolution of the heat pulse
consists in substituting the multi-time scale series  into
equations of HST (\ref{V} )-(\ref{G}). Gathering terms of the same
order of magnitude with respect to $\epsilon $ we come up with a
chain of equations leading to divergent (secular) solutions.
Cancelling step by step these secularities we then obtain a
hierarchy of equations of different orders in parameter $\epsilon
$, governing different stages of the evolution. Omitting details
of computations we concentrate on the first order in $ \epsilon
$.  In the first order, equations for normal velocity and
temperature variations read

\[
\frac{\partial V_{0}^{\prime 3}}{\partial t_{1}^{\prime
}}\;=\;\frac{
\partial ^{2}V_{0}^{\prime }}{\partial x^{\prime }{}^{2}},\;\;\;\;\frac{
\partial T^{\prime }}{\partial t^{\prime }}\;+\;\frac{\partial V_{0}^{\prime
}}{\partial x^{\prime }}=0,
\]

 Excluding quantity $V_{0}^{\prime }$ from the
set of equations written above we obtain
\begin{equation}
\frac{\partial T^{\prime }}{\partial t_{1}^{\prime }}\;=\;\epsilon
^{-1/3} \frac{\partial }{\partial x^{\prime }}\left(
\frac{\partial T^{\prime }}{
\partial x^{\prime }}\right) ^{1/3}. \label{DR}
\end{equation}

Relation (\ref{DR}) coincides with the widely used nonlinear
heat-conductivity equation derived by Dresner \cite{Dresner} by a
pure phenomenological model. We  stress, that unlike the Dresner
derivation, our work shows that this relation is just the
long-time asymptotic limit of the full set of equations of HST.

\section{MEASUREMENTS OF THE VORTEX
LINE DENSITY IN THE SECOND SOUND RESONATOR}

The second problem which we study with use of multi-scale
asymptotic perturbation techniques is  scrutinizing a very popular
experimental method:  measuring  vortex line density with a second
sound resonator. Although that method is standard and used in many
works, we, for the sake of definiteness, will consider the recent
experiments made by Hilton and Van Sciver \cite{vs}, They
performed direct measurements  of quantum turbulence induced by
second sound shock pulses in a wide channel. In Fig.1a. there is
depicted the excess attenuation of second sound in the resonator
placed across the channel, along  which the intense thermal pulse
is propagating. It was assumed that thermal pulse created the
quantum turbulence which lead to the excess attenuation of second
sound. Assuming further that the the excess attenuation $\alpha
_{v}(t)$ is connected with the amplitude of resonance wave $A(t)$
by the well-known Hall-Vinen relation $ \alpha _{v}(t)=\alpha
_{v}(0)(A(0)/A(t)-1)$ , and \ expressing \ $\alpha _{v}(t)$ via
the VLD $L(t)$ the authors drew a conclusion about the VLD
evolution, coinciding (up to factor) with the curve depicted in
Fig.1a.

\begin{figure}[tbp]
\centerline{\psfig{file=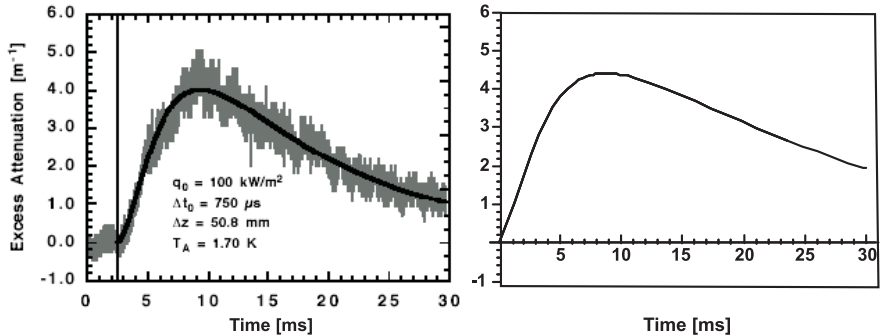,height=1.75in}}
\caption{(a) Excess attenuation of the resonance waves measured by
author of \cite{vs}, (b) The same quantity computed with use of
asymptotic perturbation techniques. {\it We have to stress that
this behavior is just the transient response of the resonator, and
change of amplitude does not follow (directly) the evolution of
VLD.} }
\end{figure}

It is easy to see, however, that this evolution looks to be quite
unreasonable. Indeed, according to current view on VLD dynamics,
the quantity $L(t)$ should grow only in presence of relative
velocity. But the pulse duration was $0.75\;ms$ , whereas $L(t)$ \
increases during $5\div 7\;ms$. The origin of that discrepancy is
obvious. The Hall-Vinen relation  based on the quality factor is
valid only for very slowly changing external conditions (duration
of thermal pulse) in comparison with the time constant of the
resonator. Here the condition was  inverted and the Hall-Vinen
relation was inapplicable; i.e. the amplitude $A(t)$ evolved
independently from the quality factor evolution. In other words
the curve depicted in Fig.1.  demonstrates  reaction of the
resonator rather than the true behavior of VLD $L(t)$.

Thus the direct procedure of measuring of VLD based on the quality
factor is not correct and the question appears whether it is
possible at all to extract information about VLD $L(t)$ evolution.
The answer is positive and corresponding analysis is based on the
general topic of our work i.e. on the fact that excess attenuation
of the second sound due to quantum turbulence changes slower than
duration of main thermal pulse. Then  we are in position to apply
the multi-scale asymptotic perturbation techniques.

Let us describe very briefly the corresponding computations.
Taking the spacial Fourier transform to Eqs. (\ref{V}) we arrive
at the following equation for variations of the temperature
$T^{\prime }$ in the standing second sound wave

\begin{equation}
\frac{d^{2}T^{\prime }}{dt^{2}}+p(\xi )\frac{dT^{\prime }}{dt}+\omega
^{2}T^{\prime }=\omega ^{2}T_{0}e^{i\omega t}  \label{NayfehEq}
\end{equation}

Here $T^{\prime }_{0}$ is the pumping amplitude, $\ \omega $ $\
$is the resonance frequency of standing wave, $\ p=\alpha
_{b}+\alpha _{v}(\xi )\ \ $is $\ \ $ the attenuation $\
$coefficient consisting of background part $ \alpha _{b}\ $\ due
to background vorticity and variable part $\ \alpha _{v}(\xi )\
$due to passing thermal pulse. The latter \ depends on the slow
time $ \xi =\epsilon t,$ where $\epsilon $ is small parameter,
characterizing, say, attenuation coefficient/frequency. The
corresponding analysis demonstrates that the solution is (as
usual) a combination of general and particular solution. The
general
solution is just natural oscillations, attenuating proportional to \\
$\exp [-(1/2)\int p(\xi )d\xi]$.  As for particular solution, we
seek it in a form:
\begin{equation}
T_{part}^{\prime }=\eta (t,\epsilon )e^{i\omega t},\;\;\;\;\;\;\;
\frac{d\eta }{dt}=[G(\xi ,\epsilon )-i\omega ]\eta +H(\xi
,\epsilon ) \label{eta_eq_1}
\end{equation}
Substituting (\ref{eta_eq_1}) into (\ref{NayfehEq})  and comparing
coefficients near $\eta (t,\epsilon )e^{i\omega t}$ and $
e^{i\omega t}$ we obtain relations for $G$ and $H$. Representing,
further, quantities $G$ and $H$ in form of series in $\epsilon $,
and equating coefficient near the same powers of $\epsilon $ we
obtain recurrent equations for successive determination of both
$G(\xi )$ and $H(\xi )$. Restricting ourseves to the first order
we obtain the following equation to determine the complex
amplitude of particular solution, which is just time dependent
amplitude in a resonator measured in experiment \cite{vs}

\begin{equation}
\frac{d\eta }{dt}=-\frac{\alpha _{b}+\alpha _{v}(\xi )}{2}\eta +(1-\frac{%
(\alpha _{b}+\alpha _{v}(\xi ))}{4i\omega })\frac{T_{0}\omega
}{2i} \label{eta_eq_2}
\end{equation}

It is seen that amplitude of second sound wave in resonator is
equal to the pumping amplitude multiplied by the quality factor
only in limit of very long time. Generally that amplitude obeys a
more complicated law.   Let us discuss  how to extract the
information concerning true behavior of vortex line density. Here
we have performed the following procedure. We supposed that the
Vinen equation for evolution of VLD can be indeed extrapolated to
very high value of applied counterflow velocity. Then we varied
coefficients $\alpha $and $\beta $ of that equation in order to
select best values to fit the curve depicted in curve of Fig.1a.
In Fig.1b. is a curve which corresponds to values $\alpha
\approx 0.13$ $\alpha _{Vi}$and $\beta \approx 0.16$ $\beta _{Vi%
}$, where  $\alpha _{Vi}$and  $\beta _{Vi}$ are values offered by
Vinen. Evolution of the VLD corresponding to these values is
depicted in Fig.2.

\vspace{0.5cm}\includegraphics[width=6.0cm,]{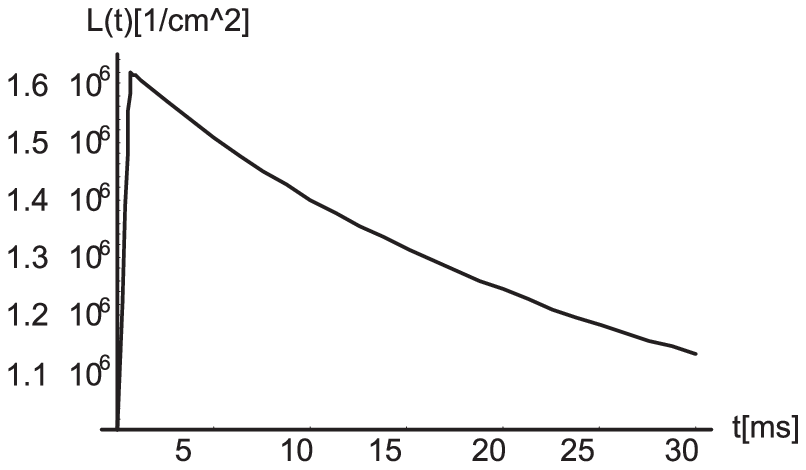}

\parbox[t]{5.8cm} { }

\vspace{-4.0cm} \hspace{6.2cm}
\parbox[t][5cm]{5.5cm}{\small Fig.2. Plausible behavior of VLD
resulting in response of resonator depicted in Fig.1. This curve
was obtained as solution of Vinen equation with parameters
obtained by fitting procedure. One can see, that VLD increases
only during thermal pulse duration $0.75\;ms$. } \vspace{-2.0cm}

\section{CONCLUSION}

By use of multi-scale perturbation analysis it has been shown how
the wave-like behavior of the heat pulse described by initially
hyperbolic equations gives way to the parabolic-type non-linear
heat conductivity equation. By the similar method we study
unsteady processes in the second sound resonator.

    Author is grateful to Hilton, D. K., Van Sciver, S. W. for useful discussion.
This work was partially supported by grant N 03-02-16179 from RFBR
and by grant N 2001-0618 from INTAS.

\end{document}